\documentclass[aps,prl,showpacs,amsmath,amssymb,amsfonts,twocolumn, superscriptaddress,floatfix,footinbib]{revtex4-1}

\usepackage[T1]{fontenc}
\usepackage{graphicx}
\usepackage{color}
\usepackage{ulem}

\newlength{\halfcolumnwidth}
\newcommand\latticeparam{\Lambda}
\newcommand\latticeparamval{830~\mathrm{nm}}
\newcommand\holeradius{r_h}
\newcommand\holeradiusval{276~\mathrm{nm}}
\newcommand\thickness{h}
\newcommand\thicknessval{213~\mathrm{nm}}

\newcommand\angleone{3.57}
\newcommand\angletwo{1.23} 
\newcommand\anglethree{0.33} 

\newcommand\affiliationLKB{Laboratoire Kastler Brossel, UPMC-Sorbonne Universit\'es,
 CNRS, ENS-PSL Research University, Coll\`ege de France, 4 place Jussieu,
 Case 74, F75252 Paris Cedex 05, France}
\newcommand\affiliationLPN{Laboratoire de Photonique et de Nanostructures LPN-CNRS/CNRS, Route de Nozay, 91460 Marcoussis, France}

\newcommand\affiliationPseven{Universit\'e Paris Diderot, Sorbonne Paris Cit\'e, 75207 Paris Cedex 13, France}

\newcommand\affiliationIBM{IBM Research - Z{\"u}rich, S{\"a}umerstrasse 4, 8803 R{\"u}schlikon, Switzerland}

\begin{document}
\title{High-finesse Fabry-Perot cavities with bidimensional Si$_3$N$_4$ photonic-crystal slabs}

\date{\today}

\author{X. Chen}
\author{C. Chardin}
\author{K. Makles}
\author{C. Ca{\"e}r}
\altaffiliation[Current affiliation: ]{\affiliationIBM}
\author{S. Chua}
\affiliation{\affiliationLKB}

\author{R. Braive}
\affiliation{\affiliationLPN}
\affiliation{\affiliationPseven}

\author{I. Robert-Philip}
\affiliation{\affiliationLPN}

\author{T. Briant}
\author{P.-F. Cohadon}
\author{A. Heidmann}
\author{T. Jacqmin}
\author{S. Del\'eglise}
\affiliation{\affiliationLKB}
\maketitle

\section*{Abstract}
{\bf
Light scattering by a two-dimensional photonic crystal slab (PCS) can result in dramatic interference effects associated with Fano resonances. Such devices offer appealing alternatives to distributed Bragg reflectors or filters for various applications such as optical wavelength and polarization filters, reflectors, semiconductor lasers, photodetectors, bio-sensors, or non-linear optical components. Suspended PCSs also find natural applications in the field of optomechanics, where the mechanical modes of a suspended slab interact via radiation pressure with the optical field of a high finesse cavity. The reflectivity and transmission properties of a defect-free suspended PCS around normal incidence can be used to couple out-of-plane mechanical modes to an optical field by integrating it in a free space cavity. Here, we demonstrate the successful implementation of a PCS reflector on a high-tensile stress Si$_3$N$_4$ nanomembrane. We illustrate the physical process underlying the high reflectivity by measuring the photonic crystal band diagram. Moreover, we introduce a clear theoretical description of the membrane scattering properties in the presence of optical losses. By embedding the PCS inside a high-finesse cavity, we fully characterize its optical properties. The spectrally, angular, and polarization resolved measurements demonstrate the wide tunability of the membrane's reflectivity, from nearly 0 to 99.9470~$\pm$ 0.0025 \%, and show that material absorption is not the main source of optical loss. Moreover, the cavity storage time demonstrated in this work exceeds the mechanical period of low-order mechanical drum modes. This so-called resolved sideband condition is a prerequisite to achieve quantum control of the mechanical resonator with light.}

\section{Introduction}

Fano resonances resulting from the interference between a discrete resonance and a broadband continuum of states are ubiquitous in physical science and constitute a unique resource in the field of nanophotonics \cite{Zhou2014}. The scattering properties of subwavelength periodic structures can be tailored by engineering their asymmetric Fano lineshapes. These extraordinary and controllable optical properties hold great promise in the context of nanophotonics, where fully integrated Photonic Crystal Slabs (PCS) can replace traditional Bragg multi-layers for various optical components. This approach was already successfully implemented in the context of optical filters \cite{Lin2005}, switches \cite{Yacomotti2006}, sensors \cite{Magnusson2011}, lasers \cite{Yang2015}, slow-light \cite{Yanik2005} and non-linear devices \cite{Yacomotti2006a}. It was also proposed as a possible route to reduce thermal noise in the mirrors of large-scale gravitational-wave interferometers \cite{Friedrich2011}.

Moreover, nanophotonics is playing an increasing role in the field of optomechanics, by allowing sub-micron-scale mechanical resonators to interact via radiation pressure with highly confined optical fields. A very successful approach consists in colocalizing a photonic and a phononic mode in the small volume of a defect cavity in the plane of a PCS \cite{Gavartin2011, Alegre2011}. In this setup, the mechanical resonator consists in a GHz internal vibrational mode localized in a very small mode volume around the defect cavity. More recently, defect-free suspended PCS have been used as lightweight mirrors of submicron thickness \cite{Kemiktarak2012b,Bui2012a,Makles2015}. The out-of-plane vibrational modes of the membrane can be coupled to an optical field by integrating the membrane inside a high finesse cavity. Because of their very low mass and low mechanical frequency in the MHz range, such resonators are highly susceptible to radiation pressure. Moreover, they are good candidates to realize hybrid opto-electro-mechanical transducers \cite{Andrews2014a, Bagci2014} as the out-of-plane vibrational modes of a suspended PCS can be simultaneously coupled to a nearby interdigitated capacitor. Such a device could play a decisive role in future (quantum) information networks as it allows to up-convert microwave signals onto an optical carrier for subsequent detection or transmission into optical fibers.

Suspended membranes made from thin layers of highly stressed materials are an advantageous platform for the development of optomechanical PCS as they present excellent mechanical properties \cite{Purdy2012a}. For instance, recent experiments have demonstrated Q-factors exceeding $10^8$ with high stress Si$_3$N$_4$ films at room \cite{Reinhardt2015a, Norte2015} and cryogenic \cite{Yuan2015} temperatures. Moreover, on the optical side, a stringent requirement is the so-called resolved sideband (or good cavity) limit \cite{Schliesser2008} in which the cavity storage time exceeds the mechanical period. To date, PCS optomechanical resonators have been demonstrated with 1D periodic patterns (often referred to as High Contrast Grating) \cite{Kemiktarak2012b,Stambaugh2014} or 2D photonic crystal patterns \cite{Makles2015, Bui2012a,Norte2015} that are insensitive to the incident polarization and that preserves the mechanical stiffness of the membrane. However, in these experiments, the PCS structures were suffering from a low optical reflectivity, precluding the operation in the resolved sideband regime and in most cases, the PCS was fabricated in a low tensile stress material, preventing from reaching high mechanical quality factors.

Here, we report on the implementation and full optical characterization of a PCS directly etched on a high tensile stress 200~nm-thick Si$_3$N$_4$ nanomembrane. We have reached the resolved sideband condition for the first time with a PCS used as the end-mirror of a Fabry-Perot cavity. The cavity finesse, of 6385 $\pm$ 150 is, to the best of our knowledge, an unprecedented value with free-space PCS cavities. The inferred membrane reflectivity is given by $1 - R = 530~\pm$ 25~ppm. Moreover, we measured the band diagram of the guided resonances responsible for the Fano resonances in the PCS with an angle resolved spectrometer. In contrast with previous works conducted in the context of PCS optical filters \cite{Qiang2008} and reflectors \cite{SinghChadha2013}, we give a 
physical interpretation of the observed band diagram and shed new light on the physics of these structures. Finally, we introduce a theoretical description of the scattering properties of the PCS in the presence of optical losses and demonstrate that the current reflectivity is not yet limited by material absorption. Demonstrating high-reflectivity PCS in a high-tensile stress film such as Si$_3$N$_4$ as evidenced here could also be beneficial in fields different from optomechanics, such as generation of tunable optical filters \cite{Stomeo2010} or optomechanical lasers \cite{Yang2015}.

\begin{figure*}
\includegraphics[scale = 0.66]{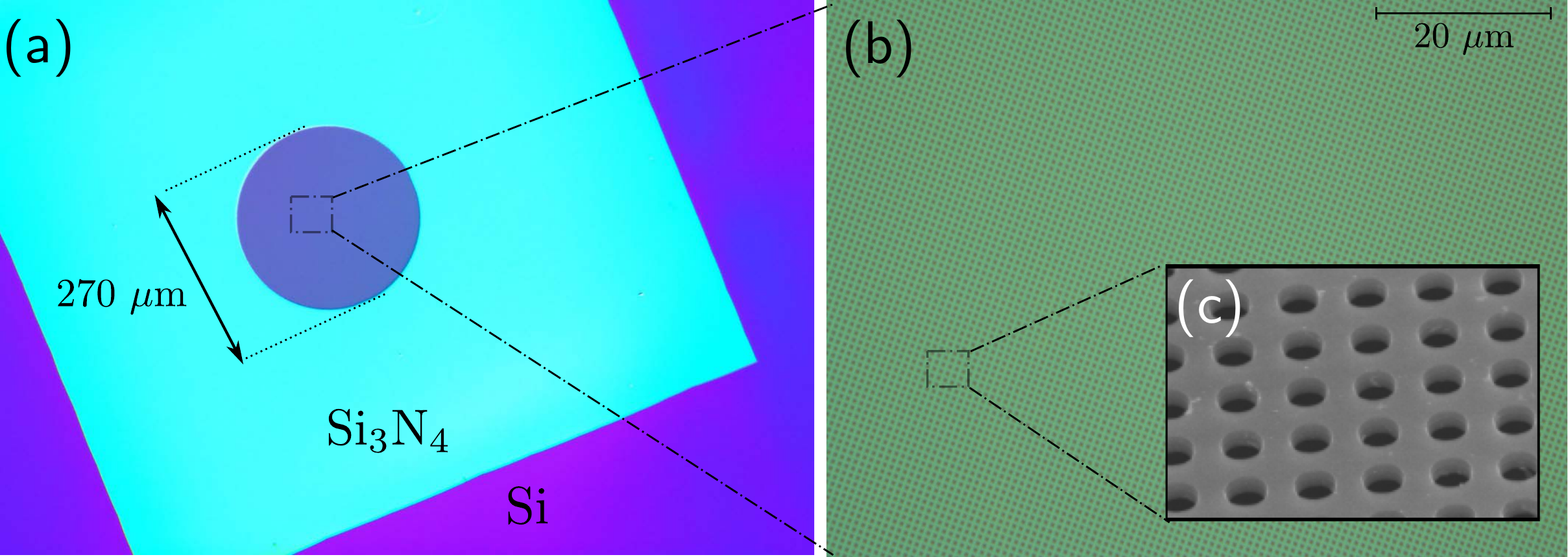}
\caption{(a) Optical image of the PCS made of a high-stress $\mathrm{Si}_3\mathrm{N}_4$ membrane suspended on a $\mathrm{Si}$ frame. The disk in the centre is the photonic crystal. (b) Optical microscope image of the PCS. (c) Scanning electron micrograph of the PCS.}
\label{Fig0}
\end{figure*}

\section{methods}

\subsection{A. Membrane and photonic crystal fabrication}

The commercial membranes~\cite{Norcada} are made of a $200$~nm stoichiometric Si$_\mathrm{3}$N$_\mathrm{4}$ film deposited on a Si substrate and released over a 1~mm$\times$1~mm square by chemical etching of the substrate (see Fig.~\ref{Fig0}a). High tensile stress in the film ensures mechanical modes with quality factors as high as $10^8$ at low temperature~\cite{Yuan2015} and ultra-low mass~\cite{Purdy2012}. The photonic crystal is realized in the centre of the membrane. It is obtained by electron-beam lithography on a layer of 200~nm of Poly(methyl methacrylate) (PMMA) resist, followed by reactive ion etching with a plasma of $\mathrm{CHF}_3$ and $\mathrm{SF}_6$. The PMMA is then stripped off with an oxygen plasma. The photonic crystal consists of a 270-$\mu$m diameter disk patterned with a square lattice of circular holes (see Fig.~\ref{Fig0}). The lattice parameter is $\latticeparam = \latticeparamval$, and the design value for the hole radius is $293$~nm.

\subsection{B. PCS characterization}
\subsubsection{Band diagram measurement}

\begin{figure*}
\includegraphics[scale = 1]{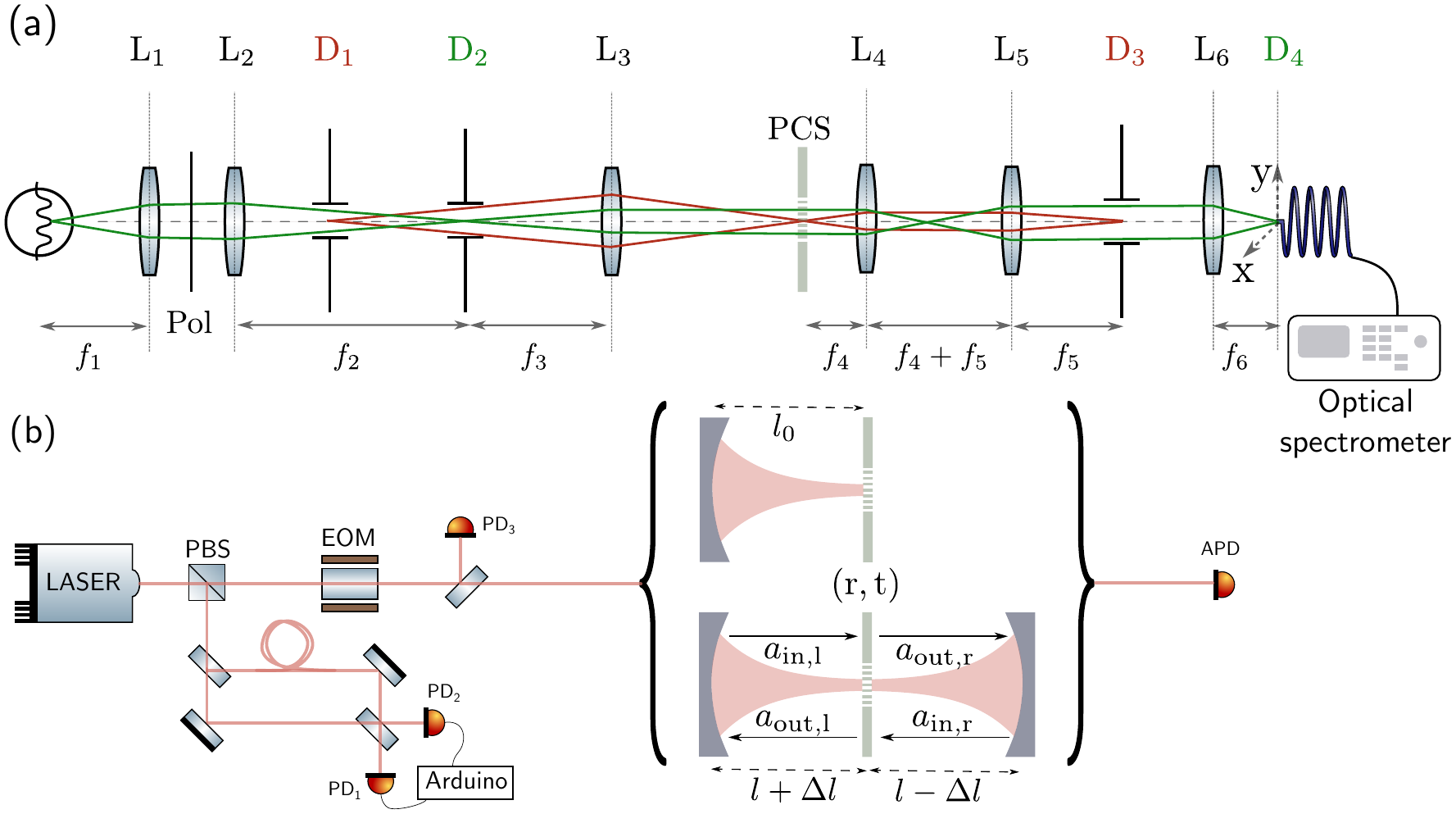}
\caption{(a) Setup used to record the PCS band structure. The light source is a Thorlabs tungsten halogen white lamp (SLS201). The illumination system is made of three lenses ($L_1$, $L_2$, $L_3$) and two diaphragms ($D_1$, $D_2$). The imaging system is made of one microscope objective ($L_4$), two lenses ($L_5$, $L_6$) and two diaphragms ($D_3$, $D_4$), ($D_4$) being the fiber tip which can be scanned along the $x$ and $y$ directions. (Pol) is a linear polarizer and (PCS) is the photonic-crystal sample. The green rays are the Fourier image rays whereas the red rays are the real image rays. (b) Fabry-Perot cavity setups used to measure the membrane reflectivity and losses:
the transmission of the cavity is measured by an avalanche photodiode (APD), and normalized  by the incident signal from the  photodiode (PD$_3$). For the single-ended cavity measurement (top cavity and Fig.~\ref{Fig3}), the laser diode is continuously scanned while an imbalanced interferometer is used to monitor its wavelength (PD$_1$, PD$_2$), and 50-MHz sidebands are generated by an electrooptic modulator (EOM). For the membrane-in-the-middle measurement (bottom cavity and Fig.~\ref{Fig4}), the laser wavelength and membrane positions are scanned by piezoelectric actuators on the grating of the Littmann-Metcalf laser system and on the membrane mount respectively. (PBS): Polarizing Beam Splitter.}
\label{Fig1}
\end{figure*}

We implemented the white-light illumination and imaging setup shown in Fig.~\ref{Fig1}(a) to measure the PCS band diagram. A halogen white lamp is combined with a set of 3 lenses ($\mathrm{L}_{1}$, $\mathrm{L}_\mathrm{2}$, $\mathrm{L}_\mathrm{3}$) and two diaphragms ($\mathrm{D}_\mathrm{1}$, $\mathrm{D}_\mathrm{2}$) to realize a K\"{o}hler illumination configuration. A field diaphragm ($\mathrm{D}_\mathrm{1}$) conjugated to the PCS restricts the illumination to  the PCS area thus reducing the impact of light scattered on optical elements and mounts. The aperture diaphragm ($\mathrm{D}_\mathrm{2}$) is used to select the incident beam angles at the PCS position. It is centred on the optical axis and its aperture width is chosen to reject high angle-of-incidence components that are irrelevant for this experiment. A magnified image of the PCS is obtained with a microscope objective ($\mathrm{L}_\mathrm{4}$) and a lens ($\mathrm{L}_\mathrm{5}$). At the image plane sits a field diaphragm ($\mathrm{D}_\mathrm{3}$), conjugated to diaphragm ($\mathrm{D}_\mathrm{1}$), providing additional filtering of unwanted scattered light. A fiber tip is positioned in the focal plane of lens ($\mathrm{L}_\mathrm{6}$), that is in the Fourier plane of this imaging system. The light collected by the fiber is guided to a near-infrared spectrometer. By scanning the fiber in the transverse (x, y) plane the transmission spectrum can be measured at different angles. Moreover, a polarizer (Pol) is introduced between ($\mathrm{L}_\mathrm{1}$) and ($\mathrm{L}_\mathrm{2}$) for polarization selection.

\subsubsection{PCS reflection, transmission and losses measurements}

We implemented a single-ended cavity with the PCS as the end-mirror (Fig.~\ref{Fig1}b) to characterize its reflectivity, transmission and losses. The input mirror of the cavity is a 20~mm radius-of-curvature mirror and the cavity length is 17.4~mm. This ensures a waist size $w_0 = 48~\mu$m located on the PCS, smaller than the lateral extent of the photonic crystal, and an  angular dispersion $\Delta \theta = \arctan\left(\lambda/\pi w_0\right) = 7.1 \times 10^{-3}$~rad that has a negligible influence on the optical properties of the PCS. A motorized Littman-Metcalff stabilized diode laser system from Sacher GMBH (Lion TEC-520) is used to probe the cavity. The cavity transmission is monitored with an avalanche photodiode (Thorlabs APD110C), and recorded continuously while the laser frequency is scanned over 4.8~nm. Simultaneously, the fringes of an imbalanced fiber interferometer are counted by a digital system (based on an Arduino micro-controller \cite{Arduino}) to measure the laser frequency in real time. The free spectral range, bandwidth and visibility of each peak are extracted from Lorentzian fits. 

\subsubsection{Description and characterization of the PCS losses}

In the following we provide a complete theoretical description of the PCS optical properties in the presence of losses. The membrane can be described by its scattering matrix, transforming the fields $a_\mathrm{in,l}, a_\mathrm{in,r}$ incoming from left and right into outgoing fields $a_\mathrm{out,l}, a_\mathrm{out,r}$ (see Fig. \ref{Fig1}b):
\begin{align}
\left(\begin{array}{cc}
a_\mathrm{out, l} \\
a_\mathrm{out, r}
\end{array} \right) 
&= S \left( \begin{array}{cc}
a_\mathrm{in, l} \\
a_\mathrm{in, r}
\end{array}
\right),\\ 
\rm{\,with\,\,\,} 
S = \left(\begin{array}{cc}
r & \bar t \\
t & \bar r \\
\end{array} \right) &= \left(\begin{array}{cc} 
r &  t \\
t &  r \\
\end{array} \right),
\end{align}
where the symmetry of the membrane with respect to the $xOy$ plane is used to assume that the direct $(r,t)$ and reciprocal $(\bar r, \bar t)$ complex reflection and transmission amplitudes are equal. 
The previous assumption allows $S$ to be diagonalized
\[
S =  V \left(\begin{array}{cc}
r + t & 0 \\
0 &  r - t \\
\end{array} \right) V^{-1},
\]
with the change of basis
\[
V =  \frac{1}{\sqrt{2}}\left(\begin{array}{cc}
1 &  1 \\
1 &  -1 \\
\end{array} \right)
\]
For a lossless membrane, $S$ is unitary ($|r + t| = |r-t| = 1$), such that within a global phase factor, it is entirely determined by a single real parameter, for instance the intensity transmission $T = |t|^2$. However, in the presence of losses, the scattered fields have a total intensity smaller than the input fields and both eigenvalues obey
\begin{align}
\label{eq:bounds}
\left|r + t\right| \leq& 1 \\
\left|r - t\right| \leq& 1.
\end{align}
Three real parameters (plus a global phase term) are required to fully describe $S$: in addition to the intensity transmission $T$, we introduce the optical losses
\begin{equation}
L_\pm = 1 - |r \pm t|^2
\end{equation}
associated to the symmetric and antisymmetric combinations of left and right incoming fields 
${\left(\begin{array}{c}
a_\mathrm{in, l}  \\
a_\mathrm{in, r}  \\
\end{array} \right) \propto \frac{1}{\sqrt{2}} \left(\begin{array}{c}
1  \\
\pm 1  \\
\end{array} \right)}$.
To obtain a measurement of $L_\pm$, the membrane can be positioned in the centre of a high-finesse cavity. This setup, initially proposed in the group of J. Harris~\cite{Thompson2008a}, consists of two single-ended cavities coupled via the transmission of the membrane. If the input and output mirrors have the same amplitude reflectivity $r'$ and if both sub-cavities have the same length $l$, one gets in the absence of fields incident from outside the cavity (see Fig. \ref{Fig1}b):
\begin{align}
a_\mathrm{in,l} &= t r' \mathrm{e}^{-2 i k l} a_\mathrm{in,r} + r r' \mathrm{e}^{-2 i k l} a_\mathrm{in,l} \\
a_\mathrm{in,r} &= t r' \mathrm{e}^{-2 i k l} a_\mathrm{in,l} + r r' \mathrm{e}^{-2 i k l} a_\mathrm{in,r},
\end{align}
where $k= \tilde \omega/c$ is the wave-number of the propagating field. 
The eigenmodes of the total optical cavity correspond to non-trivial solutions of this system for complex angular frequencies $\tilde{\omega}$. By taking the sum and difference of the previous equations, we obtain
\begin{align}
a_\mathrm{in,l} + a_\mathrm{in,r} =& (r+t) r' \mathrm{e}^{-2 i k l} \left[a_\mathrm{in, l} + a_\mathrm{in, r}\right]\\
a_\mathrm{in,l} - a_\mathrm{in,r} =& (r-t) r' \mathrm{e}^{-2 i k l}  \left[a_\mathrm{in,l} - a_\mathrm{in,r}\right].
\end{align}
Non-trivial solutions of this system correspond to symmetric (+) (respectively antisymmetric (-)) modes with angular frequency $\omega_{p,+}$ (resp. $\omega_{p,-}$) and damping (full width at half maximum) $\gamma_+$ (resp. $\gamma_-$) given by
\begin{align}
\omega_{p,\pm} =& \mathrm{Re}\left(\tilde{\omega}_{p,\pm}\right) = - \frac{c }{2 l} \left(\mathrm{arg}\left[ \left(r\pm t \right) r'\right] + 2 p \pi \right)\label{omegapm}\\
\gamma_\pm =& -2 \mathrm{Im}\left(\tilde{\omega}_{p,\pm}\right) = - \frac{c }{l} \mathrm{log}\left( \left|r\pm t \right| |r'|\right)\label{gammapm}
\end{align}
where $p$ is an integer corresponding to the longitudinal mode number. From Eq.~(\ref{omegapm}), the frequency splitting between symmetric and antisymmetric modes of identical longitudinal numbers is given by:
\begin{align}
\Delta \nu =& \frac{\omega_{p,+} - \omega_{p,-}}{2 \pi} = \frac{c}{4 \pi l} \mathrm{arg} \left( \frac{r-t}{r + t} \right)
\label{deltanu}
\end{align}
Finally, if the membrane losses $L_\pm$ and mirror losses $L' = 1 - |r'|^2$ are small, Eq.~(\ref{gammapm}) simplifies to
\begin{align*}
\gamma_\pm \approx& \frac{c }{2 l} \left(L_\pm + L'\right).
\end{align*}
Hence, the damping of the symmetric and antisymmetric modes gives direct access to the two parameters $L_+$ and $L_-$.

To perform this measurement, a 32-mm long cavity is formed by two mirrors with transmission-limited losses $L'= 455$~ppm, and radii of curvature 20~mm. The PCS is set in the centre of the cavity, on a three-axis translation stage. In order to align the cavity axis with respect to the PCS, the end mirror is also mounted on a two-axis translation stage. Finally, the PCS frame is glued on a piezoelectric stack to allow for automatic scans of its $z$ position over several free spectral ranges.

\section{Results and Discussion}
\subsection{A. Band diagram of the PCS}

\begin{figure*}
\includegraphics[width=\textwidth]{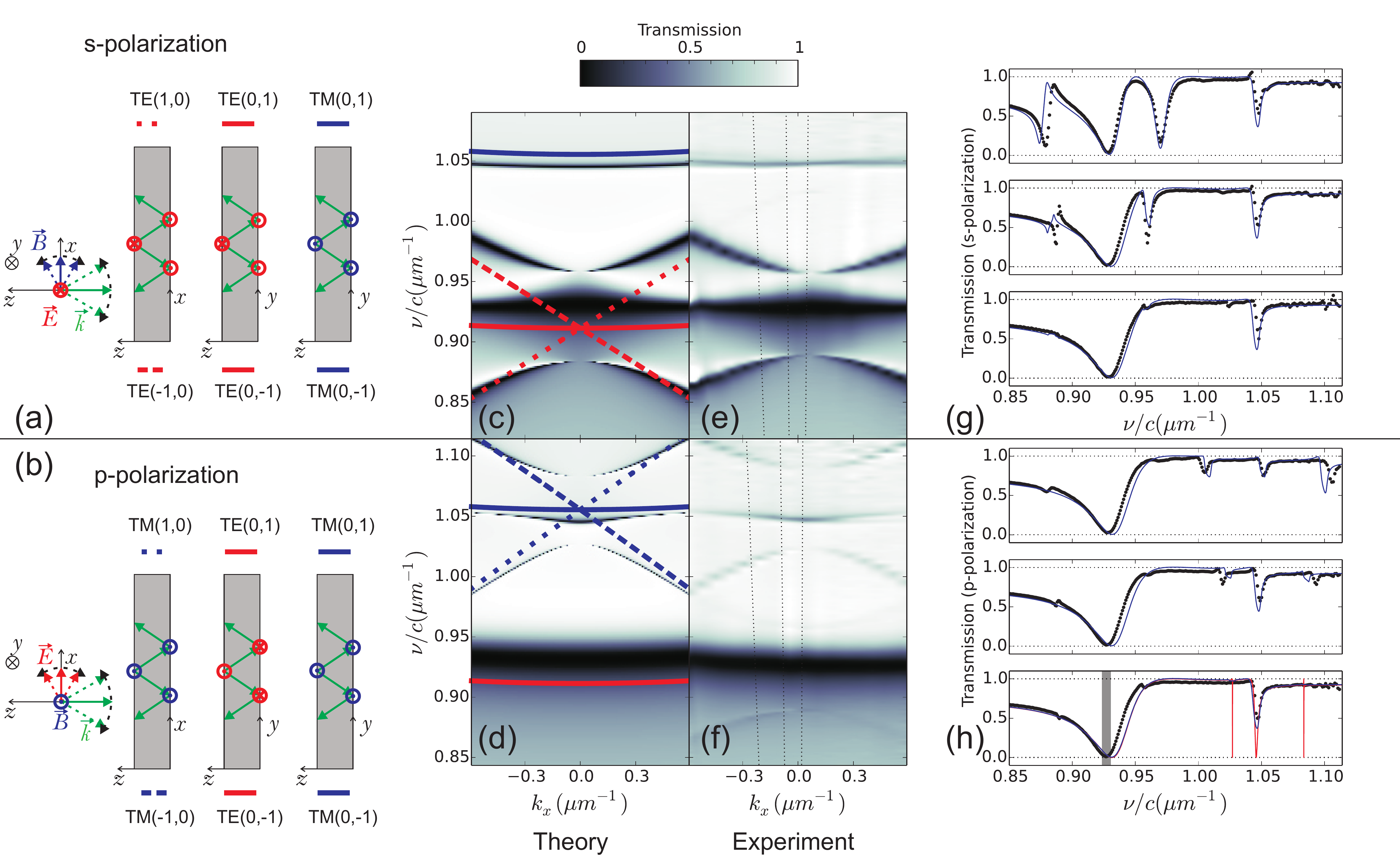}
\caption{(a),(b) First-order guided modes of an unpatterned membrane illuminated under s-polarization and  p-polarization respectively. (c), (d) Theoretical reflectivity map of a $\thicknessval$~thick PCS patterned  with a square lattice of period $\latticeparam = \latticeparamval$, round hole filling factor $\eta = 30.6~\%$, and dielectric constant given by~\cite{RI}, as a function of lateral wave-vector $k_x$, and normalized laser frequency $\nu/c$ under s-illumination (c), and p-illumination (d). The guided-resonance frequencies of the unpatterned membrane are superimposed, with TE-modes in red and TM-modes in blue (see (a) and (b) for the list of symbols). (e), (f) Experimental map of the reflectivity measured under s-illumination (e) and p-illumination (f). (g), (h) Experimental transmission (points) measured for an angle \angleone$^\circ$, \angletwo$^\circ$ and \anglethree$^\circ$ (from top to bottom) for s-illumination (g) and p-illumination (h). The dashed lines of constant $k_x/k_{z,air}$ in (e) and (f) correspond to these measurements. The corresponding theoretical RCWA fit convoluted with the spectrometer response is plotted as a blue line. As an example, the unconvoluted RCWA result is plotted in red in the last cut of (h). In this plot, the width of the shaded area corresponds to the wavelength span scanned around the TE resonance in Fig.~\ref{Fig3}.}
\label{Fig2}
\end{figure*}

The physics of high contrast gratings has previously been described in detail~\cite{Karagodsky2010,Karagodsky2012a}. 
Here, we extend it to the case of a PCS, following from earlier work ~\cite{Fan2002,Makles2015}. The high-reflection properties of the PCS result from Fano resonances arising from an interference between the direct reflection from the PCS and light leaking out from internal high-Q guided modes in the PCS. 
To unveil the band structure of the guided modes, we calculate the transmission of the PCS by means of the Rigorous Coupled Wave Analysis (RCWA)~\cite{Liu2012} which is well suited to periodic structures such as our square lattice of holes (lattice parameter $\latticeparam = \latticeparamval$, membrane thickness $\thickness = \thicknessval$, hole radius $\holeradius = \holeradiusval$). 
We begin the analysis with the unpatterned membrane limit for which the dielectric constant of both the membrane material and the holes is set to an average value $\epsilon_\mathrm{eff} = \eta \epsilon_\mathrm{air} + (1-\eta) \epsilon_{\mathrm{Si}_3\mathrm{N}_4} = 2.95$. 
The field solutions are guided modes, referred to as ``unperturbed modes'' in the following discussion. They are plane waves reflected by total internal reflection upon the two faces of the membrane. In this regime, the unperturbed modes are evanescent outside of the membrane: the $z$ component of their wavevector fulfils $k_{z,\mathrm{m}} \in \mathbb{R}^+$ inside the membrane, and $k_{z, \mathrm{air}} \in i \mathbb{R}^-$ in the surrounding medium. TE (TM) modes have their electric (magnetic) field within the plane of the membrane, and are represented on Fig.~\ref{Fig2} in red (blue) respectively. The eight modes are sorted according to their symmetry class with respect to the $xOz$ plane. Symmetric modes ($TM(\pm 1, 0 )$) are not excited under s-illumination whereas antisymmetric modes ($TE(\pm 1, 0)$) are not excited under p-illumination.
The complex coefficients describing the reflection inside the membrane for the electric field are given by Fresnel's law
\[
r_\mathrm{TM} = \frac{ k_{z,\mathrm{m}} - \epsilon_\mathrm{eff} k_{z, \mathrm{air}}}{ k_{z,\mathrm{m}} + \epsilon_\mathrm{eff} k_{z, \mathrm{air}}}  \mathrm{, \,\,\,}
r_\mathrm{TE} = \frac{k_{z, \mathrm{air}} - k_{z,\mathrm{m}}}{k_{z,\mathrm{m}} + k_{z, \mathrm{air}}}.
\]
We have numerically solved the round-trip resonance condition within the membrane
\begin{equation}
r_\mathrm{j}^2 \mathrm{e}^{- 2 i k_{z,\mathrm{m}} \thickness}  = 1,
\end{equation}
for ${j} = {TE}$ or ${TM}$, and for the lowest possible value of $\omega = c \sqrt{\left({k_x} + 2 p_x \pi/\latticeparam \right)^2 + \left(2 p_y \pi/\latticeparam \right)^2 + k_{z,\mathrm{m}}^2}$,  (where $(p_x, p_y) = (\pm 1, 0)$ or $(0, \pm 1)$ are the diffraction orders in the $x$ and $y$ directions respectively). Note that the periodicity $\latticeparam$ is not a property of the unpatterned membrane and that it is arbitrarily chosen to match that of the patterned membrane.
Given the phases of $r_\mathrm{TE}$ and $r_\mathrm{TM}$, we find that TE modes have a lower fundamental resonance frequency than TM modes. Finally, the symmetry of the mode with respect to the $xOy$ plane is given by the sign of the half round-trip propagation coefficient 
\begin{equation}
r_\mathrm{j} \mathrm{e}^{- i k_{z,\mathrm{m}} \thickness}  = \pm 1.
\end{equation}
From the expressions of $r_\mathrm{TE}$ and $r_\mathrm{TM}$, it follows that the fundamental TE (resp. TM) mode is antisymmetric (symmetric) with respect to $xOy$. The calculated mode frequencies as a function of transverse wavevector $k_x$ are represented by the lines in Fig.~\ref{Fig2}c and d. To first order, the frequency of the modes diffracted along the $y$ direction is not affected by the transverse $k_x$ component, contrary to the modes diffracted along the $x$ direction.

The results of the RCWA for the patterned membrane are superimposed as a color plot in Fig.~\ref{Fig2}c and d (refractive index of $\rm{Si}_3{N}_4$ given by \cite{RI}). 
In this case, the modes described in the unpatterned case are now coupled together by diffraction, giving rise to avoided crossings. In addition, the reflections at the membrane/air interfaces couple the guided modes to the zeroth-order modes propagating in free-space, leading to an effective loss channel. The guided resonances acquire a finite lifetime inversely proportional to the width of the Fano resonance (cut along a fixed value of $k_x$ in Fig.~\ref{Fig2}). 
We find that the lifetime of TM modes is approximately 20 times that of TE modes. 
Note that at normal incidence the $yOz$ plane constitutes an additional plane of (anti)-symmetry in the s- (p-) polarization, and hence, only the combinations $TE(1, 0) + TE(-1, 0)$ and $TM(0,1) + TM(0, -1)$ $\left(\mathrm{resp.\,\,} TE(0, 1) + TE(0, -1) \mathrm{\,\,and\,\,} TM(1, 0) + TM(-1, 0)\right)$ are effectively coupled to free space modes: the linewidth of the other Fano resonances vanish close to $k_x=0$. The white light angle-resolved spectrometer described previously allowed us to measure the band structure of guided modes for small transverse wave-vectors $k_x$. 
The experimental transmissions with s- (p-) illumination is presented in Fig.~\ref{Fig2}e and \ref{Fig2}f, together with 3 different cuts with incidence angle $\angleone^\circ, \angletwo^\circ$ and $\anglethree^\circ$. The simulation is fitted to the data with the hole radius and membrane thickness as free parameters. Indeed, the former is only specified to within 10 \% accuracy and the latter may differ from the design value due to imperfect pattern transfer upon electron-beam lithography and etching. 

The fitted values $\holeradius = \holeradiusval$ and $\thickness=\thicknessval$ are in good agreement with the design values. The theoretical fits are represented in blue in Fig.~\ref{Fig2}g and h. The reduced visibility of the peaks, particularly pronounced with TM modes, is well accounted for by convolving the theoretical map with the 4-nm wide response of the spectrometer. In the next section, we analyse in more detail the TE Fano resonance responsible for the largest dip in transmission close to $\nu = c \times 0.929 \,\mu \mathrm{m}^{-1} = c / (1.076 \,\mu \mathrm{m})$ in Fig. \ref{Fig2}g and h.

\begin{figure*}
\includegraphics[width=1.05\textwidth]{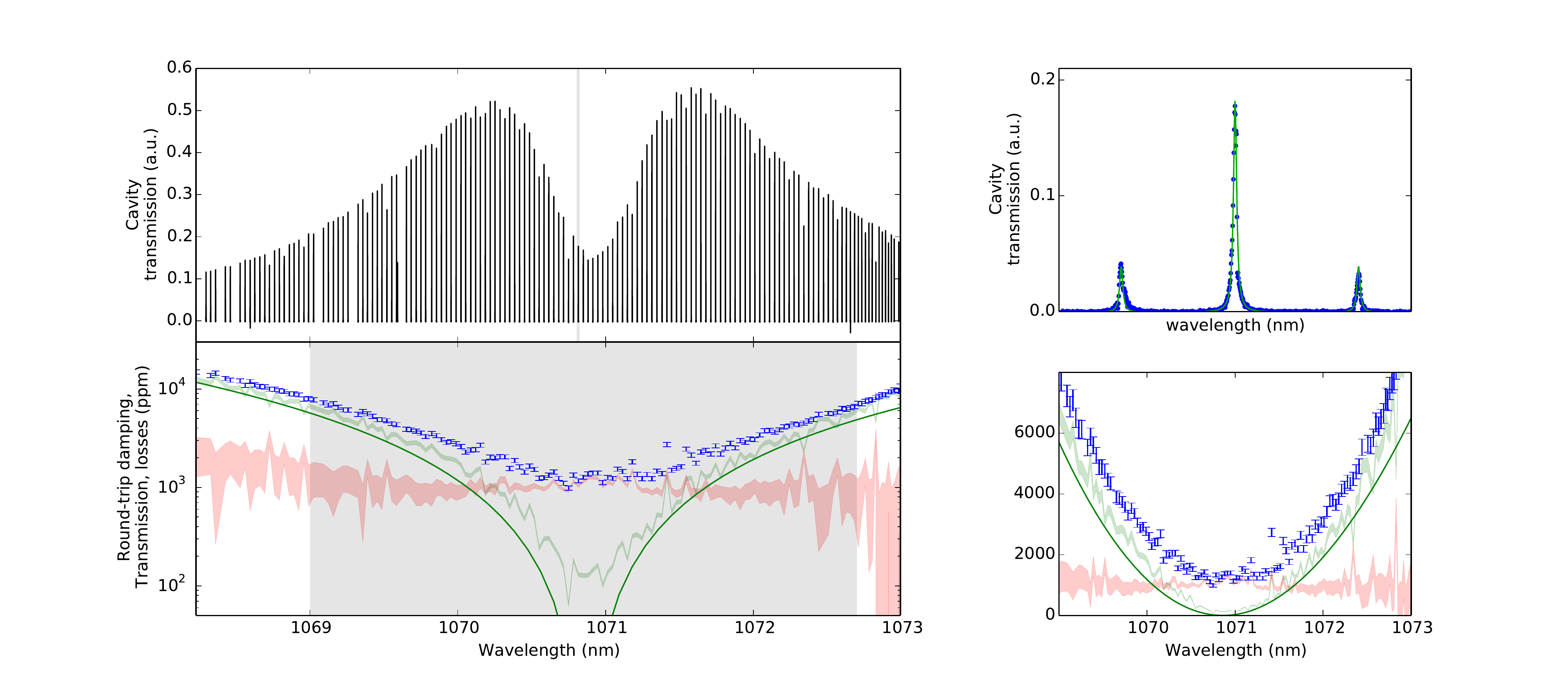}
\caption{(a) Transmission of a single-ended cavity as a function of laser wavelength. (b) Zoom of an individual Airy peak, showing modulation sidebands, and a Lorentzian fit. (c) Round-trip loss inferred from the fits of individual Airy peaks in blue. The error bars are extracted from the average off-centering of the central peak between left and right sidebands. The red and green shaded area represents the confidence interval for the PCS loss and transmission respectively. The green line is the theoretical transmission calculated by RCWA with the same parameters as Fig. \ref{Fig2}, except for $r_h = 285$~nm (this experiment uses a second membrane from the same micro-fabrication run). (d) provides a linear representation of the data in (c).}
\label{Fig3}
\end{figure*}
\subsection{B. Detailed characterization of a Fano resonance}
For optical frequencies $\nu=c/\lambda$ close to a guided mode resonance, the PCS complex reflection and transmission coefficients $r$ and $t$ are given in good approximation by the simple two-mode model~\cite{Fan2002}
\begin{align}
r(\lambda) =& r_d \pm  \frac{\gamma}{i (\lambda - \lambda_0) + \gamma} ( \pm r_d - t_d)\label{fano1}\\
t(\lambda) =& t_d + \frac{\gamma}{i (\lambda - \lambda_0) + \gamma} (\pm r_d - t_d),
\label{fano2}
\end{align}
where $r_d$ and $t_d$ are the off-resonant coefficients (${|r_d + t_d|}={|r_d - t_d|}=1$), $\lambda_0$ is the wavelength of the guided mode resonance, $\gamma$ the resonance width, and the plus/minus sign in Eq.~(\ref{fano1}) reflects the symmetry of the guided mode: as discussed in the previous section, the lowest order TE-mode is antisymmetric with respect to the $xOy$ plane, and thus the minus sign applies.
 In the absence of optical losses, a striking feature of this Fano lineshape is the existence of a frequency $\lambda_1 = \lambda_0 + i \gamma r_d/t_d$ for which the PCS transmission vanishes:  $T(\lambda_1) = |t(\lambda_1)|^2 = 0$, and accordingly, the reflectivity tends to unity : $R(\lambda_1) = |r(\lambda_1)|^2 = 1$.

By fitting the RCWA results with Eqs.~\ref{fano1} and \ref{fano2} around the largest TE Fano resonance, we determine $\lambda_1 = 1076$~nm and $\gamma = 12$~nm. In order to study the optical behaviour of the PCS around $\lambda_1$, a single-ended optical cavity of length $l_0$ was realized, using the PCS as end-mirror (see Fig. \ref{Fig1}). The transmission of the cavity is monitored while a tunable diode laser is swept across the Fano resonance. The  recorded signal, shown in Fig.~\ref{Fig3}a, displays a comb of peaks spaced by the free spectral range of the cavity given by
\begin{equation}
\Delta \lambda_\mathrm{FSR} = \frac{\lambda_0^2}{2 l_0}.
\end{equation}
The linewidth $\Delta \lambda_n(\lambda)$ of each longitudinal cavity mode $n$ is then obtained by fitting a Lorentzian profile to each of the individual peaks. To minimize the impact of scan speed fluctuations, we generate 50-MHz sidebands in the laser spectrum with an electro-optic modulator, used as a frequency calibration of the instantaneous scanning speed. Denoting the finesse of the cavity by $\mathcal{F(\lambda)}$, the total round-trip loss of the cavity is given by $\Gamma_\mathrm{RTL}(\lambda) = 2 \pi/\mathcal{F(\lambda)} = 2 \pi \Delta \lambda_n(\lambda) / \Delta \lambda_\mathrm{FSR}(\lambda)$. They are plotted in blue in Fig.~\ref{Fig3}b as a function of the laser wavelength. 
The lowest level of $\Gamma_\mathrm{RTL} = (985 \pm 25)$~ppm, corresponding to a finesse $\mathcal{F}(\lambda_1) =  6385 \pm 150$, was found at $\lambda_1=1070.9\,$nm. The $5.1$~nm discrepancy with respect to the results of Fig.~\ref{Fig2} is quite small considering that two different membranes from the same micro-fabrication run were used in these experiments. In the simulations, it is accounted for by changing the hole radius from $r_h = 276$~nm to $r_h = 285$~nm. Taking into account the 455 ppm transmission of the input mirror, we infer a membrane reflectivity as good as $1 - R(\lambda_1)$ = 530~ppm.

To discriminate between the contributions of the membrane transmission $T(\lambda) = |t(\lambda)|^2$ and the optical losses $L(\lambda) = 1 - R(\lambda) - T(\lambda)$ in $\Gamma_\mathrm{RTL}(\lambda)$, we use the extra-information embedded in the height of the transmitted peaks. Indeed, the transmission $\mathcal{T}_n(\lambda)$ of the cavity at resonance with the $\mathrm{n}^\mathrm{th}$ mode is given by
\begin{equation}
\label{trans}
\mathcal{T}_n(\lambda) = \frac{4 T_c T(\lambda)}{\Gamma_\mathrm{RTL}(\lambda)^2},
\end{equation}
where $T_c$ is the transmission of the input mirror, assumed to be constant over the wavelength range scanned here. In the experiment, a signal proportional to $\mathcal{T}_n(\lambda)$ is measured, so that we can deduce $T(\lambda)$ from Eq.~(\ref{trans}) within a multiplicative constant $\alpha$. 
By assuming that the losses $L(\lambda)$ are within the interval $[0, \Gamma_\mathrm{RTL}(\lambda_1)]$ at the highest accessible wavelength, we obtain a confidence interval for $\alpha$, and hence for $T(\lambda) = \Gamma_\mathrm{RTL}(\lambda) - L(\lambda)$ (in green on Fig.~\ref{Fig3}c and d) 
and for $L(\lambda)$ (in red) over the whole wavelength range. For comparison, the transmission calculated by RCWA is displayed with a green solid line on Fig.~\ref{Fig3}c and d. The agreement between RCWA and experimental results is excellent. At the Fano resonance $\lambda_1 = 1070.9$~nm, the round-trip damping is vastly dominated by membrane losses. The lowest cavity bandwidth $\kappa/2 = c/4 l \mathcal{F}(\lambda_1) = 675$ kHz measured in this single-ended cavity setup approaches the resolved-sideband condition for first-order flexure modes of the suspended membrane.

\begin{figure*}
\centering
\includegraphics[height=0.49\paperheight]{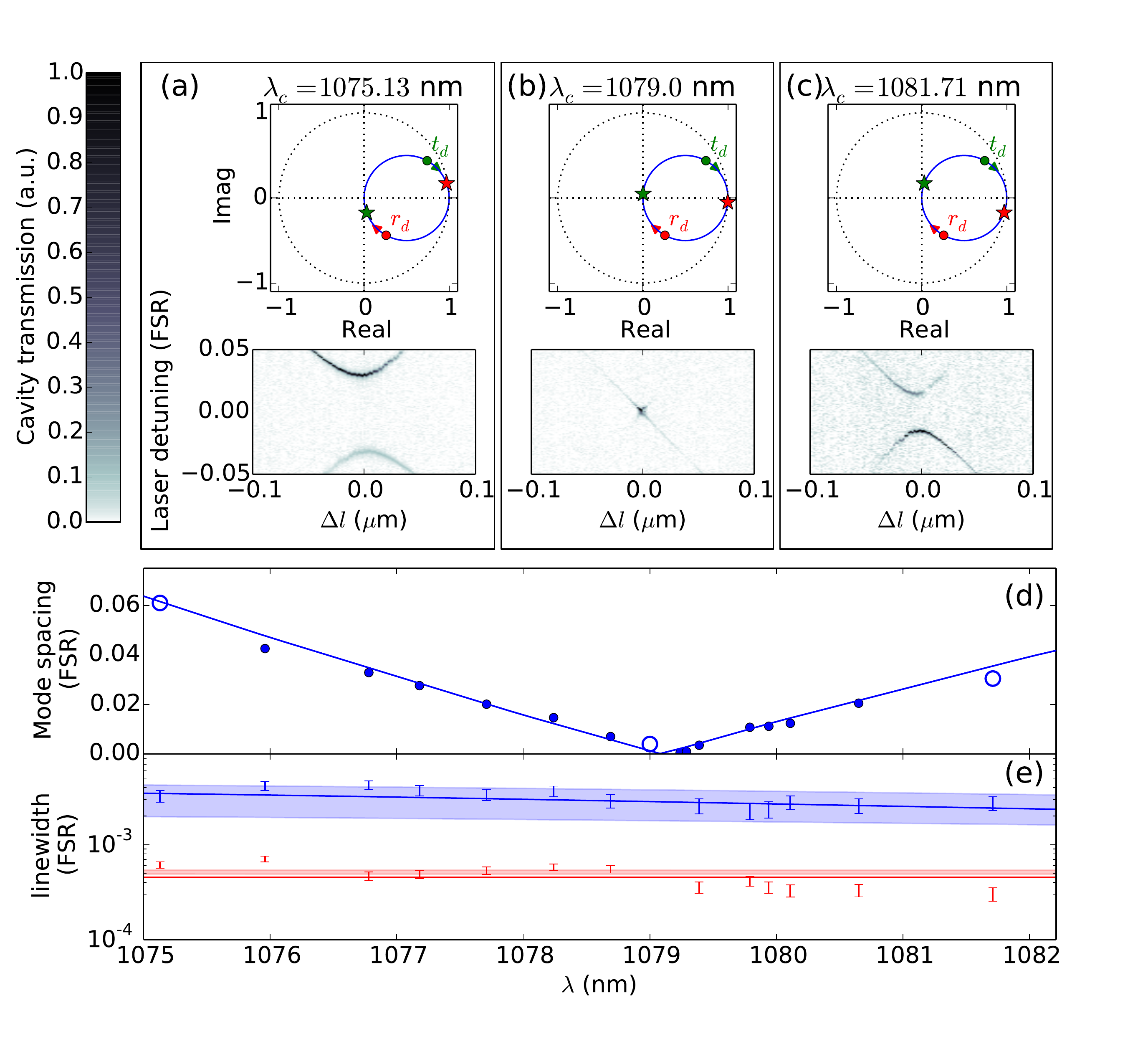}
\caption{(a), (b) and (c) show colorplots of the cavity transmission versus the cavity length and laser detuning with respect to three different laser central wavelengths $\lambda_c$. Above are the corresponding complex representations of $r(\lambda)$ and $t(\lambda)$, where a global phase factor has been applied to ensure that $r(\lambda_1) = r_d+t_d=1$. The green and red bullets correspond to $r(\lambda_0) = t_d$ and $t(\lambda_0) = r_d$ respectively, the green and red stars correspond to $t(\lambda_c)$ and $r(\lambda_c)$ respectively. In (a) ((c)), $\lambda_c$ is chosen smaller (larger) than $\lambda_1$. In (b), $\lambda_c$ is chosen very close to $\lambda_1$, so that $r(\lambda_c)\simeq 1$ and $t(\lambda_c)\simeq 0$. (d) Mode frequency spacing for $\Delta l = 0$. A minimum is reached for $\lambda = \lambda_1$. The blue line corresponds to the mode spacing calculated by RCWA using Eq.~(\ref{deltanu}) (same parameters as in Fig.~\ref{Fig2} and \ref{Fig3}, except for $r_h = 272$~nm). (e) Linewidths of the two modes. The blue strip represents the value obtained with RCWA simulations including an absorption of $2 \times 10^{-5}\leq \mathrm{Im}(n)\leq 5 \times 10^{-5}$. Optical detunings are normalized by the FSR of the sub-cavities $\Delta \lambda_\mathrm{FSR} = \lambda_c^2/2 l$. The error bars are inferred from the variation of the fit parameters obtained with neighbouring values of $\Delta l$.}
\label{Fig4}
\end{figure*}

To understand the exact form of the scattering matrix, and gain insight on the origin of the losses observed in the previous experiment, a PCS is positioned in the middle of a high-finesse cavity. The PCS position $\Delta l$ is scanned along the cavity axis with a piezoelectric actuator. In this configuration, the PCS couples two cavities of length $l_1 + \Delta l/2$, and $l_2 - \Delta l/2$, where $l_1 \approx l_2 \approx l$ are the initial lengths of the first and second sub-cavities, supposed to be simultaneously resonant with the central laser wavelength $\lambda_c$ for $\Delta l = 0$. The cavity transmission, recorded as a function of $\Delta l$ and laser detuning $\delta \nu \propto \lambda_c - \lambda$, is represented in the color plots of Fig.~\ref{Fig4}a, b and c. The avoided crossing of symmetric and antisymmetric modes is visible in the centre of the figure. From a Lorentzian fit of the peaks at the avoided crossing and along the vertical axis, we extract the frequency spacing $\Delta \nu$, and the linewidth $\gamma_\pm$ of symmetric and antisymmetric modes. The experiment is then reproduced for several laser wavelengths $\lambda_c$ close to the Fano resonance and the corresponding frequency spacings and linewidths are shown in Fig.~\ref{Fig4}d and e respectively.

In Fig.~\ref{Fig4}a and c, the  
upper and lower branches of the avoided crossing are strongly asymmetric. 
This is a direct consequence of the difference between $L_+$ and $L_-$. For $\lambda_c<\lambda_1$ (Fig.~\ref{Fig4}a), the phase of $(r-t)/(r+t)$ is positive, such that from Eq.~(\ref{deltanu}), the symmetric mode (angular frequency $\omega_+$) constitutes the upper branch of the crossing, while the antisymmetric mode (angular frequency $\omega_-$) is the lower branch. On the other hand, for $\lambda_c > \lambda_1$ (Fig.~\ref{Fig4}c), $\Delta \nu = (\omega_+ - \omega_-)/2\pi$ changes its sign and the upper and lower branches are exchanged. Finally, at the Fano resonance $\lambda_c = \lambda_1$, $r$ and $t$ are in phase, so that symmetric and antisymmetric modes are perfectly degenerate. Combined with the inequalities of Eq.~(\ref{eq:bounds}), an upper bound for the transmission of the membrane can be derived: $|t| \leq 1 - |r|$ at the Fano resonance. For $1 - R = 985$~ppm, this corresponds to a transmission as low as $T \leq 0.2$~ppm. Hence, the residual transmission of 100 ppm observed in Fig.~\ref{Fig3} is somehow in contradiction with the two-port description developed in this work. A first explanation is the possible scattering into non-overlapping spatial modes due to surface roughness. Moreover, given our experimental parameters, we expect from RCWA that the wavevector components of the input Gaussian beam not aligned with a symmetry plane of the membrane should give rise to $\approx 20$ ppm transmission into the polarization orthogonal to the input beam (with a spatial wavefunction that is antisymmetric with respect to the $xOz$ and $yOz$ planes). Nevertheless, as visible on Fig.~\ref{Fig3}, the residual transmission only accounts for a small fraction of the total cavity losses close to the Fano resonance, and the nature of these losses still needs to be elucidated.

The difference between the losses $L_\pm$ experienced by the symmetric and antisymmetric modes is well captured by a simple phenomenological model: since the Fano resonance results from the interference between the direct PCS transmission and a high-Q guided resonance, it can be safely assumed that the losses mostly affect the former. Hence, Eqs.~(\ref{fano1}) and (\ref{fano2}) are modified to include a phenomenological loss rate $\gamma'$ to the high-Q response as
\begin{align*}
r =& r_d - \frac{\gamma}{i (\lambda - \lambda_0) + \gamma + \gamma'} (-r_d - t_d) \\
t =& t_d + \frac{\gamma}{i (\lambda - \lambda_0) + \gamma + \gamma'} (-r_d - t_d).
\end{align*}

Using this model, it follows that 
\begin{align}
L_+ &= 0 \label{asymlosses}\\
L_- &= \frac{4 \gamma \gamma'}{(\gamma + \gamma')^2 + (\lambda - \lambda_0)^2} \label{symlosses},
\end{align}
and thus, only the antisymmetric mode of the two-cavity setup is affected by the optical losses, with a value proportional to the Lorentzian response of the guided resonance. This is an intuitive result since the antisymmetric guided resonance under study isn't driven by a symmetric incoming field (this would be vice-versa for the symmetric TM resonance). To quantitatively evaluate the model, Fig.~\ref{Fig4}e shows the total expected linewidths $L_\pm + L'$, where $\gamma' = 1.0 \times 10^{-2}$~nm is the only adjustable parameter, since $\lambda_1 = 1079.1$~nm, $\gamma = 12$~nm and $L' = 455$~ppm are independently determined. Therefore, optical losses are found to account for less than $0.1~\%$ in the total decay rate of the guided mode resonance. The previous model shows that the asymmetry between $L_+$ and $L_-$ is a direct consequence of the physics of Fano resonances, and that it is not specific to a particular loss mechanism such as absorption or scattering, since both would preferentially affect the high-Q guided mode. This finding is in stark contradiction with a model recently proposed in~\cite{Stambaugh2014}, where such an asymmetry was attributed to absorption in the material. Moreover, we show in the following that absorption should have a negligible influence on the optical loss of our PCS. Indeed, if the losses observed in this experiment were entirely due to optical absorption, the value $\mathrm{Im}(n)= (3.5 \pm 1.5) \times 10^{-5}$ could be derived for the absorption of the $\mathrm{Si}_3\mathrm{N}_4$ layer. The values of $L_\pm + L'$ obtained with a RCWA simulation including an absorption $2 \times 10^{-5} \leq \mathrm{Im}(n) \leq 5 \times 10^{-5}$ are located within the shaded areas in Fig.~\ref{Fig4}e. Since previous experiments~\cite{Serra2015a} have found values for the absorption of a similar layer an order of magnitude smaller  ($\mathrm{Im}(n) = 2.0\times 10^{-6}$), we conclude that scattering is still the main source of losses. In the single-ended cavity case where the field is essentially incident on one side of the membrane, the relevant optical losses are given by $L = \left(L_+ + L_- \right)/2$. Thus, the decrease of optical losses with optical wavelength visible in Fig.~\ref{Fig3}b is consistent with Eq.~\ref{asymlosses} and \ref{symlosses}. With the previous smaller value of $\mathrm{Si}_3\mathrm{N}_4$ absorption, a RCWA simulation shows that a reflectivity as high  as $R \approx 1 - 80$ ppm, corresponding to a finesse close to 80 000, could be achieved by improving the PCS fabrication process. 

\section{Conclusions}
High-stress $\mathrm{Si}_3\mathrm{N}_4$ films have been successfully patterned to realize a 2D photonic crystal slab. We have provided a detailed study of the rich mode structure underlying the high-reflectivity of the crystal and have fully characterized the scattering and loss properties of the device. 
We have also experimentally demonstrated the wide tunability of the membrane reflectivity, via the adjustment of the laser wavelength close to a Fano resonance.  Such tunability is highly desirable in the context of membrane-in-the-middle optomechanics experiments. In particular, the ability to reach very narrow energy gap between the eigenmodes of the coupled cavities could be used to enter the regime where the photon exchange time becomes comparable to the mechanical period. This would enable the observation of mechanically-induced coherent photon dynamics, like Autler-Townes splitting or Landau-Zener-Stueckelberg dynamics~\cite{Heinrich2010}. 
Finally, the bandwidth of the single-ended cavity described in this work, as low as 675 kHz sets us in the resolved sideband regime for the first order harmonics of the drum mode, ranging from 596 kHz for the (0, 1) mode to 1.2 MHz for the (2, 2) mode used in previous quantum optomechanics experiments \cite{Purdy2012a}. Hence, this technique results in devices already suitable for the development of optomechanical transducers.

\section*{Acknowledgements}
The authors gratefully acknowledge M. Corcos, I. Krasnokutska, X. Pennehouat and L. Pinol for preliminary experimental developments. We thank J. Palomo and M. Rosticher for developing the first steps of the microfabrication process. Finally, we thank J. Lawall and R. Gu\'erout for insightful scientific discussions. This research has been partially funded by the Agence Nationale de la Recherche programs ``ANR-2011-BS04-029 MiNOToRe'' and ``ANR-14-CE26-0002 QuNaT'', the Marie Curie Initial Training Network ``cQOM'', and the DIM nano-K Ile-de-France program ``NanoMecAtom''. S. C. is supported by the Marie Sklodowska-Curie Individual Fellowship program.

\bibliography{biblio.bib}

\end{document}